# In-Situ Timing Diagnosis of PDN and Configuration-Upset-Induced Routing Delay Degradation in SRAM-based FPGAs

Mostafa Darvishi, *Senior Member, IEEE*

**Abstract**— Timing degradation in SRAM-based FPGAs arises from multiple physical mechanisms that manifest differently in the routing fabric, most notably power-distribution-network (PDN) marginality and configuration-induced routing perturbations. Existing in-situ timing monitors provide limited insight into the physical origin, spatial structure, or statistical characteristics of the degradation. This paper presents a scalable in-situ timing diagnosis architecture that enables fine-grained, routing-aware characterization of timing behavior directly within the FPGA fabric during normal operation. The proposed approach combines non-intrusive delay taps placed at routing switch-matrix boundaries with distributed phase-swept delay monitoring elements and centralized statistical analysis. By extracting probabilistic delay distributions rather than binary timing margins, the framework captures both mean delay shifts and timing variability across spatially distributed routing locations. Experimental results obtained on a modern SRAM-based FPGA show that PDN-induced timing degradation produces globally correlated delay shifts with minimal change in variance, whereas routing-induced perturbations exhibit localized, topology-dependent delay growth and increased timing dispersion. Spatial correlation analysis and two-dimensional correlation heatmaps further reveal distinct signatures that enable systematic differentiation between these mechanisms.

The presented architecture operates concurrently with an active user design and does not require external instrumentation, radiation sources, or design modification. These results establish a practical foundation for in-situ timing diagnosis, reliability assessment, and architecture-aware timing management in large FPGA-based systems.

## I. INTRODUCTION

TIMING predictability has become a dominant challenge in modern digital systems as interconnection delay increasingly outweighs logic delay and variability sources continue to intensify. In SRAM-based FPGAs, this challenge is exacerbated by the programmable routing fabric, where signals traverse multiple layers of switch matrices, programmable interconnect points, and long interconnect segments whose electrical characteristics are highly topology-dependent [1]. As a result, timing behavior in FPGAs is shaped not only by nominal design constraints but also by dynamic operating conditions and configuration state [2]-[4].

Among the most practically significant contributors to timing degradation in SRAM-based FPGAs are power-distribution-network (PDN) marginality and configuration-induced routing perturbations [5], [6]. PDN-related effects arise from transient or sustained voltage droop caused by switching activity [3], leading to correlated delay shifts across broad regions of the fabric [7], [8]. In contrast, configuration-induced routing perturbations—such as those caused by single-event upsets (SEUs) in routing configuration memory, introduce unintended parasitic connections that locally increase routing delay in a cumulative and topology-dependent manner [9]-[11]. Although both mechanisms ultimately manifest as delay increases, their physical origins, spatial extent, and temporal behavior differ fundamentally [4].

Conventional timing analysis techniques, including static timing analysis (STA) [12] and post-implementation margining [13], are ill-suited to distinguish between these degradation mechanisms once a design is deployed [14], [15]. Existing in-situ timing monitors and critical-path sensors can detect timing violations or slack erosion, but they typically provide binary or aggregate indicators that obscure the underlying cause. Consequently, designers lack visibility into whether observed timing degradation reflects a global operating condition, such as PDN stress [16], or a localized routing anomaly that may require targeted mitigation.

This paper addresses this gap by introducing an in-situ timing diagnosis framework that directly observes routing-induced timing behavior within the FPGA fabric while the user design remains fully operational. Rather than relying on functional failure or worst-case timing margins, the proposed approach extracts statistical timing distributions at physically meaningful routing locations using non-intrusive observation points and phase-swept sampling. By combining distributed sensing with centralized analysis, the framework enables spatially resolved timing characterization and systematic differentiation between globally correlated and localized timing degradation mechanisms. The practical feasibility of the proposed in-situ timing diagnosis framework is supported by a quantitative implementation analysis, with FPGA resource utilization and scalability characteristics.

Mostafa Darvishi is with Electrical Engineering Department of École de technologie supérieure (ÉTS), Montreal, Canada. He is also VP of Engineering at Evolution Optiks R&D Inc. (e-mail: darvishi@ieee.org).

The remainder of this paper is organized as follows. Section II reviews background and related work in FPGA timing analysis and in-situ monitoring. Section III discusses the physical mechanisms of timing degradation in SRAM-based FPGAs. Section IV presents the proposed in-situ timing diagnosis architecture. Section V describes the experimental methodology. Section VI reports experimental results, including baseline characterization, mechanism differentiation, and spatial correlation analysis. Section VII discusses design implications, and Section VIII concludes the paper and outlines future directions.

## II. BACKGROUND AND RELATED WORKS

Timing analysis and monitoring in SRAM-based FPGAs have been extensively studied, yet most existing approaches focus on detection rather than diagnosis [1], [9], [12], [14], [15], [17]. Static timing analysis remains the primary tool for design-time verification, relying on analytical delay models derived from characterization of routing resources and programmable switches [1], [10], [11], [17], [18]. While STA is effective for nominal timing closure, it provides limited insight into dynamic delay behavior arising from operating conditions, variability, or configuration state after deployment [19]-[21].

A range of in-situ timing monitoring techniques has been proposed to complement static analysis [6], [14], [22], [23]. These include critical-path replicas, tunable delay lines, ring oscillators, and adaptive sampling-based monitors. Such techniques have been used to detect timing margin erosion, estimate slack, or support dynamic voltage and frequency scaling. However, most monitors are designed to observe a small number of critical paths or provide aggregate indicators, making them poorly suited for capturing the spatial structure of routing-induced delay behavior [5], [9], [10], [12]-[14], [16], [17], [22], [24], [25].

Some works have also examined the impact of PDN effects on FPGA timing [2], [3], [5]-[7], [13], [16]-[22], [25], [26], demonstrating that voltage droop can induce correlated delay shifts across large regions of the fabric. Separately, radiation effects on SRAM-based FPGAs have been shown to induce routing delay changes through configuration upsets, leading to cumulative delay growth and potential timing failures. These studies, however, typically rely on external radiation facilities, fault injection frameworks, or offline analysis, and do not provide a unified in-fabric mechanism for observing and distinguishing these effects during normal operation.

More recently, statistical and spatial analysis techniques have been explored in the context of FPGA reliability and variability, but they are often applied to synthetic structures such as oscillators or test paths rather than functional routing networks [27]-[32]. As a result, the relationship between measured timing behavior and actual routed signal paths remains indirect.

In contrast to existing works, this paper introduces an architecture that anchors timing observation directly to switch-matrix traversal along functional routes, extracting statistical delay information at multiple physically meaningful locations. By combining phase-swept sampling, spatially distributed monitors, and correlation-based analysis, the proposed framework moves beyond detection toward mechanism-level diagnosis of timing degradation in deployed FPGA designs.

Table I. Comparison of In-Situ FPGA Timing Diagnosis Architectures.

| Feature/Metric | This Work | Ref. [14] | Ref. [22] |
|---|---|---|---|
| Measurement Principle | Phase-swept BER profiling | Canary / critical-path monitor | Ring-oscillator sensors |
| Delay Resolution | Sub-cycle (phase-step limited) | Binary (pass/fail) | Indirect (frequency shift) |
| Spatial Resolution | Switch-matrix level | Path-level only | Region-level |
| Intrusiveness | Non-intrusive (buffered fan-out taps) | Intrusive (path insertion) | Region-level |
| Diagnostic Capability | Distinguishes PDN vs. routing perturbations | Intrusive (path insertion) | Non-intrusive |
| Statistical Characterization | Yes (BER, CDF, σ) | No | Limited |
| Spatial Correlation Analysis | Yes (correlation & heatmaps) | No | No |
| Routing Awareness | Explicit (switch-matrix taps) | No | No |
| Hardware Overhead | Moderate, scalable | Low | Low |
| External Instrumentation | None | None | None |

To further contextualize the proposed framework relative to prior FPGA timing monitors, Table I compares this work against representative in-situ approaches reported in [14] and [22]. The comparison highlights key differences in measurement resolution, intrusiveness, spatial observability, and diagnostic capability. Unlike prior monitors that primarily provide binary timing margin or global variation indicators, the proposed architecture enables probabilistic delay extraction and explicit differentiation between PDN-induced and routing-induced timing degradation with switch-matrix-level spatial resolution.

This comparison underscores that the proposed system advances the state-of-the-art by combining fine-grained routing awareness with statistical timing analysis, enabling diagnostic capabilities that are not supported by existing in-situ monitoring techniques.

## III. TIMING DEGRADATION MECHANISMS IN SRAM-BASED FPGAS

Timing degradation in SRAM-based FPGAs can arise from multiple physical mechanisms that manifest differently at the circuit and interconnection levels [12], [15], [26]. Two of the most prominent and practically relevant sources are power-distribution-network (PDN) marginality [5], [16], [20], [22] and configuration upset-induced (SEU like) affecting routing resources [9], [10], [33], [34]. Although both mechanisms ultimately appear as path-delay increases in static timing analysis, their underlying causes, spatial characteristics, and temporal behavior differ fundamentally.

Fig. 1 conceptually illustrates the two dominant timing degradation mechanisms addressed in this work and emphasizes their distinct physical, spatial, and temporal characteristics. In the PDN-induced case, elevated switching activity increases instantaneous current demand on the power distribution network (PDN), resulting in transient supply voltage droop. This reduction in effective supply voltage degrades transistor drive strength across multiple logic blocks (LBs) and interconnect resources that share the affected power domain. As depicted in Fig. 1, the resulting delay increase evolves smoothly over time and exhibits strong spatial correlation, impacting a broad set of timing paths simultaneously. The delay magnitude is stress-dependent and largely reversible once switching activity subsides and voltage recovers [3], [4], [5], [7], [13], [20], [21].

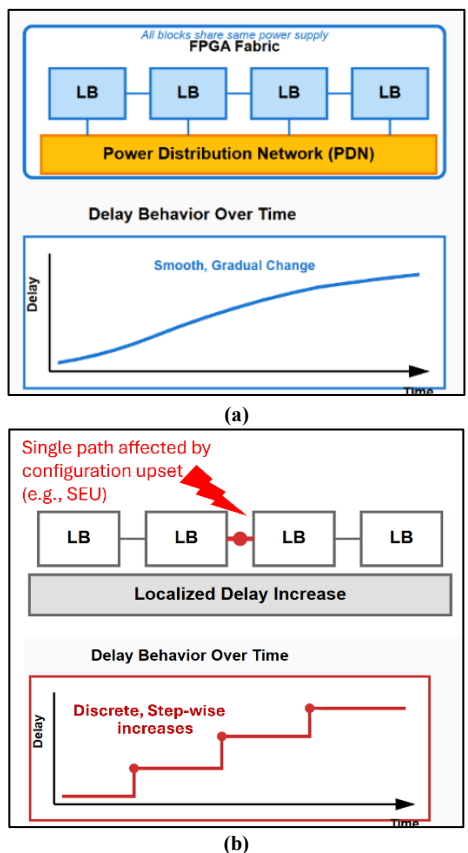

Fig. 1. Conceptual illustration of timing degradation mechanisms in SRAM-based FPGAs. PDN-induced degradation, (a), arises from transient supply voltage droop caused by increased switching activity, producing spatially correlated and smoothly varying delay shifts across multiple logic blocks. SEU-induced degradation, (b), results from configuration upset-induced in routing resources, leading to localized, discrete, and cumulative interconnect delay increases that persist until reconfiguration [20], [34].

In contrast, SEU-induced timing degradation originates from either configuration upset-induced in configuration memory bits that control programmable routing resources leading to bitflip [9]-[11], [13], [32]-[34]. Such upsets introduce unintended parasitic connections within the interconnection network of the FPGA fabric, increasing the effective resistance and capacitance of specific nets. As shown in Fig. 1, these routing perturbations produce discrete and localized delay increments that accumulate over time when multiple upsets affect the same signal path, resulting in cumulative delay change (CDC) growth. Unlike PDN-induced effects, SEU-induced delay changes are topology-dependent, spatially isolated, and persistent until the affected configuration state is corrected through reconfiguration. The distinct delay–time signatures illustrated in Fig. 1 motivate the development of an in-situ diagnosis framework capable of distinguishing timing degradation by physical origin rather than by magnitude alone [1], [5], [14], [15], [17], [21], [22], [26].

Although both mechanisms increase path delay, their timing signatures differ. PDN-induced degradation produces broadly correlated, statistically smooth delay shifts across many paths, whereas SEU-induced routing changes produce discrete, topology-dependent delay increments concentrated on specific interconnect structures. Conventional in-situ timing monitors generally lack the resolution or observability required to distinguish between these effects, motivating the need for a diagnosis framework that can separate timing degradation by physical cause rather than merely detecting its presence.

## IV. PROPOSED IN-SITU TIMING DIAGNOSIS ARCHITECTURE

This section presents the proposed in-situ timing diagnosis architecture for SRAM-based FPGAs. The architecture is designed to observe routing-induced timing behavior directly within the fabric while the user design remains fully operational. By combining non-intrusive signal tapping at routing switch matrices with distributed statistical timing monitors and centralized control, the proposed system enables fine-grained characterization of timing degradation mechanisms, including power-distribution–induced delay shifts and configuration-upset–induced routing perturbations.

Fig. 2 provides a high-level view of the architecture and its integration with the functional routing fabric. The upper portion of the figure represents the functional path of the user design, composed of logic blocks (LBs) interconnected through the FPGA routing infrastructure. Each LB is implemented within a slice, and slices reside inside configurable logic blocks (CLBs). Functional signals propagate between LBs through a hierarchy of programmable interconnect resources rather than through direct point-to-point wiring.

In Fig. 2, this routing fabric is abstracted as a sequence of switch-matrix blocks (SBs) interposed between neighboring LBs. Each SB represents the collection of programmable routing structures that a signal traverses when crossing from one logic region to another. Physically, an SB encapsulates the combined effect of local dedicated interconnects, planar switch matrices associated with CLB tiles, and larger switch matrices located in interconnect tiles. Connectivity within an SB is established through configuration-controlled programmable interconnect points (PIPs), which select specific fan-in and fan-out paths and thus define both the functional connectivity and the electrical characteristics of the route.

While Fig. 2 emphasizes the routing-centric instrumentation architecture, the monitored signals originate from a realistic functional design under test (DUT) rather than isolated test paths. In this work, the DUT is a streaming finite impulse response (FIR) filter that exercises both arithmetic and routing resources typical of signal-processing workloads on SRAM-based FPGAs. The DUT is integrated directly into the routing fabric shown in Fig. 2, and delay taps observe selected functional nets as they traverse multiple switch-matrix stages. The functional placement and routing of the DUT are held fixed across all experiments to ensure that observed timing variations arise from controlled perturbations and operating conditions rather than changes in the functional implementation. Additional controlled routing segments are instantiated alongside DUT-derived nets to systematically vary switch-matrix traversal depth and fan-out while preserving functional correctness. Detailed DUT configuration, resource utilization, and experimental conditions are described in Section V.A.

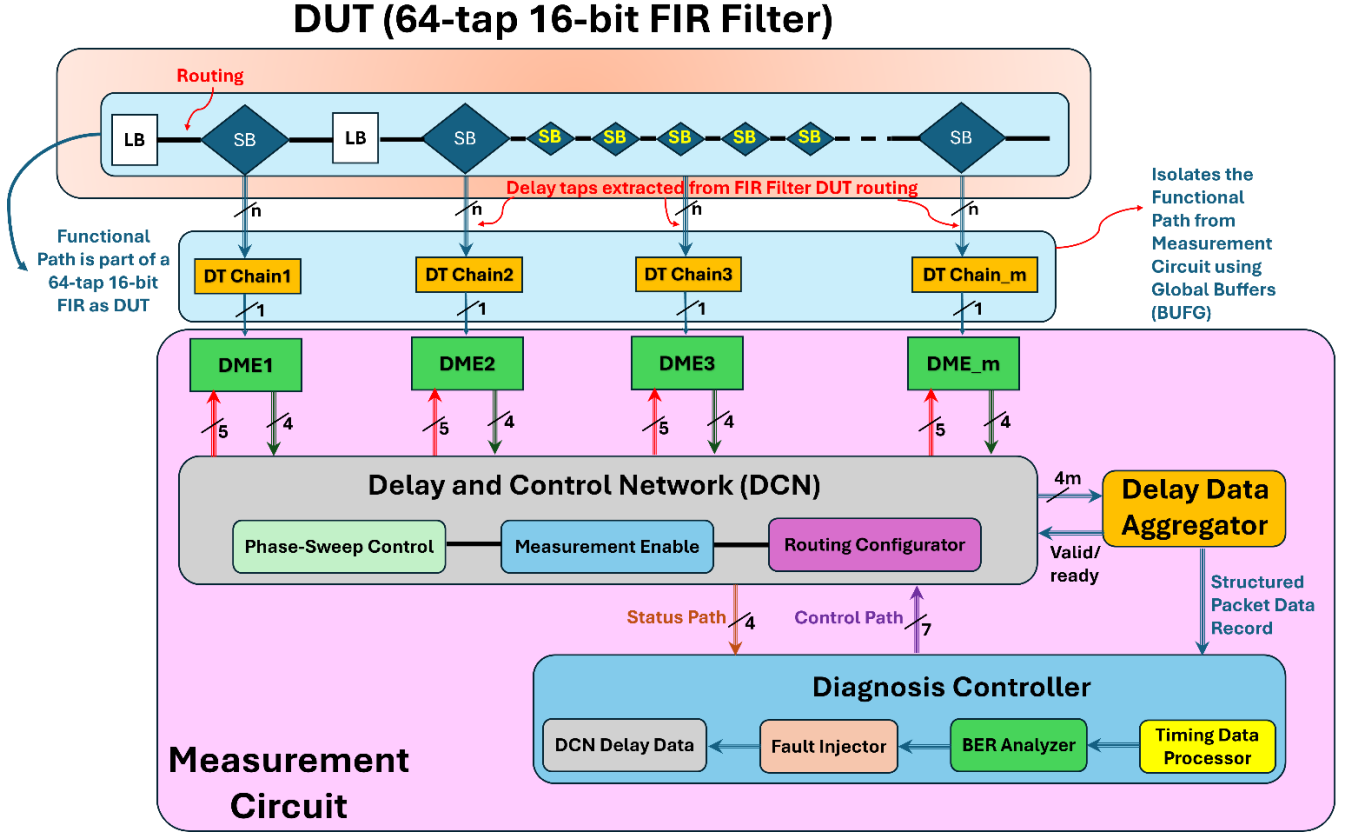

Fig. 2. High-level architecture of the proposed in-situ timing diagnosis system integrated with the FPGA routing fabric and a realistic functional design under test (DUT). Functional signals originating from the DUT propagate between logic blocks (LBs) through a sequence of switch-matrix blocks (SBs). Large SB symbols denote major switch matrices at logic block boundaries, while smaller SB symbols represent intermediate programmable interconnect points (PIPs) within routing channels that contribute incremental delay through configuration-controlled multiplexers and interconnect segments. Delay taps (DTs) non-intrusively extract buffered replicas of routed signals at selected switch-matrix I/O nodes and forward them to distributed delay monitoring elements (DMEs). The delay and control network (DCN) coordinates phase-sweep control, measurement enable, and routing-configuration commands, while the delay data aggregator and diagnosis controller collect and analyze statistical timing information across multiple DMEs. Global clock buffers (BUFGs) isolate the functional routing paths from the measurement circuitry to preserve functional timing integrity.

The number and arrangement of SBs encountered along a routed path depend on the relative placement of the source and destination LBs. Short connections may traverse only local interconnect and a planar switch matrix, whereas longer connections propagate across multiple interconnect tiles and therefore cross several switch matrices in sequence. Each SB traversal contributes incremental delay arising from wire parasitics, switch resistance, and capacitive loading, making SBs the dominant determinants of routing delay and its variability.

It is important to clarify that the routing-induced delay effects examined in this work do not rely on physical fault injection or radiation-based upset generation. Instead, the proposed architecture leverages the intrinsic programmability of the FPGA routing fabric to emulate routing-level delay perturbations in a controlled and repeatable manner. By selectively enabling additional configuration-controlled fan-out branches and parasitic routing attachments at switch-matrix nodes, the system reproduces the electrical consequences of configuration-induced routing perturbations, namely increased resistance, capacitance, and effective traversal depth, while preserving the functional routing topology of the user design. This approach allows routing-induced timing behavior to be isolated and analyzed without requiring external fault-injection infrastructure or disruptive experimental conditions.

This routing abstraction motivates the placement of delay taps at switch-matrix I/O nodes along the functional path. Following the abstraction of the routing fabric, the next layer of the proposed architecture introduces delay tap (DT) chains, which form the interface between the functional interconnect and the timing diagnosis circuitry. DT chains are designed to observe routing behavior at physically meaningful locations while preserving the electrical and timing integrity of the user design.

Each switch matrix along a routed path exposes multiple input and output nodes through which functional signals propagate. These nodes inherently support fan-out through configuration-controlled programmable interconnect points. The proposed architecture leverages this property by enabling additional, observation-only fan-out branches at selected switch-matrix I/O nodes. A DT is therefore not a new routing resource, but a configuration-controlled branch taken from an existing routing node. When enabled, this branch extracts a replica of the routed signal without modifying the original functional connection.

Because routing paths may traverse different combinations of planar and interconnect-region switch matrices depending on placement, multiple candidate tap points exist along any given functional route, each corresponding to a different physical delay point. To capture this spatial diversity, the architecture supports multiple DTs per monitored region. Although the number of possible tap locations is theoretically large, a bounded and practical subset is instantiated to control overhead. In the implemented system, eight DTs are deployed per monitored region, each corresponding to a distinct switch-matrix I/O node along representative routing path.

The DTs associated with a given region collectively form a DT chain, whose outputs are routed toward the measurement circuitry. Selection among DTs is performed through configuration bits that control which switch-matrix fan-out branches are active during a given measurement interval. This mechanism allows the diagnosis system to time-multiplex observations across multiple routing nodes while maintaining a fixed functional implementation. Over successive measurement windows, different DTs can be activated, enabling spatially resolved timing characterization without requiring re-synthesis or re-routing of the user design.

To ensure that the observation path remains electrically isolated from the functional routing, each DT output is buffered before being forwarded to the next stage of the architecture. In the proposed system, buffering is implemented using a global buffer primitive, which provides a well-defined drive strength and prevents additional capacitive loading from being reflected back onto the functional net. This isolation guarantees that enabling or disabling DTs does not perturb the timing behavior being measured. The buffered DT outputs are then delivered to the delay monitoring elements (DMEs), which form the distributed statistical sensing layer of the architecture.

Each DME constitutes a passive sensing unit responsible for converting observed signal transitions into statistically meaningful timing information. A DME observes only one DT-selected signal at a time, with selection performed through an internal input multiplexer controlled by the configuration infrastructure. The core operation of the DME is based on phase-swept sampling, in which the selected signal is sampled using a clock whose phase is systematically shifted relative to the functional clock. For a given phase setting, the DME repeatedly samples the signal over a fixed observation window and accumulates the outcomes of these comparisons, forming a statistical representation of timing behavior.

The statistical output of each DME is expressed in terms of bit-error-rate (BER) information, obtained by counting the number of incorrect samples observed during the measurement window. As the sampling phase is swept across a predefined range, the DME produces a BER-versus-phase profile that reflects the "*temporal distribution*" of signal transitions at the observed routing node. Importantly, the DME performs no local interpretation of this profile and instead reports raw statistical summaries to the control infrastructure.

It is important to note that the bit-error-rate (BER) metric used in this work does *not* represent functional correctness relative to an architectural or algorithmic reference. Instead, BER is employed as a statistical measure of *temporal uncertainty* in signal transitions with respect to a swept sampling phase. For each monitored signal, the expected value is implicitly defined by the transition timing consistency across repeated observations, rather than by knowledge of the signal's logical semantics. As the sampling clock phase is swept across the signal's transition region, metastability and setup/hold margin violations manifest as probabilistic sampling outcomes, producing the observed BER profile. Consequently, the proposed methodology does not require prior knowledge of the functional behavior of the monitored net and is applicable to arbitrary toggling signals, including datapath, control, or pseudo-random activity, provided sufficient transition density is present.

Moreover, the idea that timing impact of PDN stress is spatially/temporally continuous and timing impact of routing perturbation is "not", might seem fairly intuitive in the first glance. However, it is worth mentioning that the contribution of this work lies in demonstrating that such distinctions can be quantitatively observed and diagnosed in situ, using an embedded, non-intrusive instrumentation framework without external measurement equipment. The ability to extract spatially correlated delay signatures during normal operation enables practical diagnosis and validation of reliability mechanisms in FPGA fabrics, which is difficult to achieve with conventional post-silicon analysis methods.

All DME activity is orchestrated by the delay and control network (DCN), which provides global synchronization, phase-sweep coordination, and scalable data collection across the FPGA fabric. The DCN distributes phase indices that determine the relative sampling-clock phase applied to each DME and asserts measurement-enable signals that define the precise boundaries of each observation window. Phase updates and configuration changes are applied only at well-defined transitions between measurement windows, ensuring that timing measurements performed at different spatial locations are temporally aligned and directly comparable.

In addition to synchronizing phase-swept measurements, the DCN coordinates the collection of results from a large number of distributed DMEs. Rather than forwarding measurement outputs directly to the diagnosis controller, DMEs report low-bandwidth statistical summaries to the DCN upon completion of each observation window. The DCN arbitrates among reporting DMEs and serializes their outputs, preserving the association between each result and its corresponding DME, phase index, and configuration state while avoiding contention on global routing resources.

These generated outputs are consolidated by the delay data aggregator, which serves as the convergence point between distributed sensing and centralized analysis. The aggregator collects finalized measurement summaries from all active DMEs, buffers them as needed, and formats them into structured records suitable for system-level interpretation. By handling only low-rate statistical data, the aggregation process avoids interaction with timing-critical functional paths.

Within the measurement hierarchy, the diagnosis controller serves as the centralized coordination and analysis unit of the proposed architecture. The controller orchestrates measurement campaigns, manages configuration sequencing, and interprets aggregated timing data to reconstruct phase-dependent timing profiles and delay distributions. By correlating results across spatial locations and across repeated measurement cycles, the controller enables systematic identification of globally correlated timing shifts and localized routing-induced delay perturbations.

Through the integration of routing-aware observation points, distributed statistical sensing, coordinated control, and centralized analysis, the proposed architecture enables in-situ timing diagnosis with minimal intrusion. By anchoring timing measurements directly to switch-matrix traversal along functional routes, the architecture provides a scalable and physically meaningful framework for characterizing timing degradation mechanisms in SRAM-based FPGAs. This architectural foundation forms the basis for the experimental methodology described in the next section.

### A. Implementation-Level Realization of Timing Instrumentation

All components of the proposed timing diagnosis framework are fully synthesizable and were implemented using standard FPGA logic resources. This subsection 'summarizes' the concrete realization of the delay taps, delay monitoring elements, and phase-sweeping circuitry to clarify implementation feasibility and design constraints.

Each Delay Tap (DT) is implemented as a configuration-controlled fan-out branch originating from an existing routed net at a switch-matrix output. The tapped signal is buffered using a BUFG primitive to ensure electrical isolation from the functional path and to provide a well-defined drive strength. No additional combinational or sequential logic is inserted on the functional net itself, preserving original placement and routing.

Delay Monitoring Elements (DMEs) are implemented as synchronous digital sampling blocks operating in a dedicated instrumentation clock domain. Each DME consists of an input multiplexer, a phase-controlled sampling flip-flop, and accumulation counters. The sampling clock is derived from the system clock through programmable phase selection logic, implemented using clock multiplexers and delay elements available in the FPGA fabric. For each phase setting, the DME counts correct and incorrect samples over a fixed observation window, producing a bit-error-rate (BER) estimate associated with that phase.

Phase sweeping is realized by stepping a phase index register that controls the sampling clock selection for each DME. Phase updates occur only at observation-window boundaries to avoid metastability or partial measurements. All counters, control registers, and accumulation logic are synchronous and free of combinational feedback paths, ensuring robust timing closure.

The delay and control network, aggregation logic, and diagnosis controller are implemented using standard LUTs, flip-flops, and block RAM resources. No custom analog circuits, asynchronous logic, or vendor-specific black-box IP cores are required. All instrumentation blocks were synthesized, placed, and routed together with the functional design using standard Vivado flows, confirming that the proposed architecture is fully realizable on commercial FPGA platforms.

### B. Delay Reference, Measurement Interpretation, and Calibration

The proposed framework does not attempt to measure *absolute propagation delay* using a dedicated reference path or analog time base. Instead, timing behavior is inferred *statistically* by observing the probability distribution of signal transitions relative to a phase-swept sampling clock derived from the functional clock.

For each monitored routing node, the DME samples the observed signal using a clock whose phase is incrementally shifted across a predefined range. At early phase offsets, samples occur before the signal transition and consistently capture the previous logic value. At late phase offsets, samples occur after the transition and consistently capture the new logic value. In the intermediate region, sampling occurs near the transition edge, producing a probabilistic mix of correct and incorrect samples. The resulting bit-error-rate (BER) versus phase profile therefore encodes the relative timing distribution of the observed signal transition.

The implicit *timing reference* for all measurements is the functional *clock edge* that launches the observed transition. Because both the functional logic and the sampling circuitry are driven by clocks derived from the same source, relative alignment between measurement points is preserved across the design. No inter-DME synchronization is required beyond phase index coordination.

Calibration of the measurement framework is performed implicitly through *baseline characterization*. For each delay tap, a reference BER-versus-phase profile is first acquired under nominal operating conditions. Subsequent measurements under PDN stress or routing perturbation are then compared against this baseline profile. Timing degradation is quantified as a *relative shift* in the transition region, extracted through changes in the mean and variance of the BER distribution. This differential approach eliminates sensitivity to fixed offsets, clock skew, or static routing delay and enables consistent comparison across taps and spatial locations.

Table II. FPGA resource utilization of the proposed in-situ timing diagnosis framework implemented on the XCZU7EV device. Resource usage is reported for individual instrumentation components as well as for a representative deployment with 32 distributed DMEs. Reported values are post-synthesis results obtained from Vivado and reflect incremental overhead introduced by the instrumentation. Shared infrastructure components (DCN, aggregator, controller) are instantiated once, while DT and DME costs scale linearly with the number of monitored routing locations.

| Component | LUT | FlipFlop | BUFG | BRAM | DSP | % of FPGA |
|---|---|---|---|---|---|---|
| Single Delay Tap (DT) | 12 | 8 | 1 | 0 | 0 | <0.01% |
| Single Delay Monitoring Element (DME) | 85 | 96 | 0 | 0 | 0 | <0.05% |
| Delay & Control Network (DCN) | 420 | 380 | 0 | 0 | 0 | <0.2% |
| Delay Data Aggregator | 310 | 290 | 0 | 0 | 0 | <0.15% |
| Diagnosis Controller | 520 | 610 | 0 | 1 | 0 | <0.3% |
| FIR DUT (64-tap, 16-bit) | 8,505 | 10,200 | 0 | 0 | 64 | <2.7% |
| Total (32 DMEs + DCN + Controller) | 3,982 | 4,360 | 32 | 1 | 0 | ≈1.4% |

## V. EXPERIMENTAL METHODOLOGY

This section describes the experimental methodology used to implement, configure, and evaluate the proposed in-situ timing diagnosis architecture. The objective of the methodology is to ensure that the timing behavior observed by the diagnosis system reflects intrinsic properties of the FPGA routing fabric under controlled operating conditions, rather than artifacts introduced by design modifications or measurement intrusion. To this end, the experimental flow emphasizes reproducibility, isolation between functional and measurement paths, and systematic control of timing perturbations.

All experiments are conducted using an unmodified functional design that executes concurrently with the timing diagnosis infrastructure. The functional logic is synthesized, placed, and routed once, after which the diagnosis architecture is integrated in a manner that preserves the original routing of the user design. Delay taps, delay monitoring elements, and control logic are configured to operate alongside the functional fabric without altering its timing behavior. This approach ensures that timing data collected during experiments accurately represents the native routing characteristics of the deployed design.

The experimental methodology is organized around three core components: configuration of the diagnosis infrastructure, execution of synchronized phase-sweep measurements, and collection and interpretation of statistical timing data. For each experiment, the delay and control network coordinates the selection of delay taps, phase-sweep parameters, and routing-configuration states, while the diagnosis controller enforces a deterministic measurement schedule. Measurement data are accumulated over fixed observation windows and aggregated into structured records that preserve spatial and configuration context.

To enable meaningful comparison across different routing locations and experimental conditions, all measurements are performed using consistent phase-sweep ranges, observation window lengths, and control sequences. Environmental and operational conditions are held constant throughout each experiment, and multiple measurement cycles are executed to assess repeatability and statistical stability. The following subsections detail the implementation platform, measurement configuration, and data collection procedures used in this work.

### A. Implementation Platform and Design Integration

The proposed in-situ timing diagnosis architecture is implemented and evaluated on a contemporary SRAM-based FPGA platform representative of modern programmable fabrics. The experimental platform is based on the AMD/Xilinx XCZU7EV device from the Zynq® UltraScale+™ MPSoC family, which provides a hierarchical routing architecture, fine-grained configuration control, and rich clocking resources suitable for in-fabric timing characterization. All experiments are conducted on a ZCU104 evaluation board, which serves solely as a hosting platform and does not influence the routing or timing characteristics under study.

Table II summarizes the FPGA resource utilization of the proposed in-situ timing diagnosis framework as implemented on the XCZU7EV device. The results show that individual instrumentation components incur modest overhead, with a single Delay Monitoring Element (DME) consuming fewer than 100 LUTs and flip-flops, and delay taps limited primarily to buffering resources. For a representative deployment comprising 32 distributed DMEs, the total instrumentation occupies approximately 3,982 LUTs and 4,360 flip-flops, corresponding to less than 1.4% of the available programmable logic resources on the device. The use of BUFG resources is confined to delay tap interfaces and scales proportionally with the number of monitored routing paths, while no DSP resources are required by the framework and BRAM usage is limited to a single block associated with the diagnosis controller.

Importantly, resource usage scales approximately linearly with the number of deployed DMEs and delay taps, as shared infrastructure components such as the delay and control network and the diagnosis controller are instantiated once and remain constant. This linear scaling behavior confirms that the architecture can be expanded to larger monitored regions without introducing centralized bottlenecks or excessive routing congestion. When compared to the functional design under test—a 64-tap, 16-bit FIR filter occupying approximately 8,500 LUTs, 10,200 flip-flops, and 64 DSP blocks—the instrumentation overhead is comparable in logic usage and negligible in terms of routing complexity and timing impact. These quantitative results substantiate the scalability and non-intrusiveness claims made throughout the paper.

The architecture exhibits linear scalability with respect to the number of deployed DMEs. Each additional DME introduces a fixed incremental cost, while shared components such as the delay and control network and the diagnosis controller remain constant. When compared to the functional design under test, a 64-tap, 16-bit FIR filter occupying approximately 8,500 LUTs, 10,200 flip-flops, and 64 DSP blocks, the instrumentation overhead is comparable in logic usage and negligible in terms of routing congestion and timing impact.

The functional user design is implemented as a standalone logic subsystem and synthesized using a standard vendor-supported tool flow, i.e., AMD/Xilinx Vivado. To preserve the integrity of the routing fabric under test, all perturbation conditions are applied only to observation-side routing attachments, and the functional netlist connectivity and placement remain unchanged across all measurement conditions. No manual routing constraints, retiming directives, or timing-driven optimizations are applied after integration of the diagnosis infrastructure. This ensures that the observed timing behavior reflects intrinsic routing characteristics rather than tool-induced artifacts.

Functional designs used during characterization were selected to reflect realistic signal activity and routing utilization while allowing controlled observation of routing-induced timing behavior. Specifically, the experiments employed a streaming finite impulse response (FIR) filter as the primary design under test (DUT), configured with 64 taps and 16-bit input data and coefficients, operating at one sample per clock cycle. The FIR implementation exercises both arithmetic and routing resources and is representative of signal-processing workloads commonly deployed on SRAM-based FPGAs. In the evaluated configurations, the DUT utilized approximately 4k LUTs and 5.1k flip-flops, with 32 DSP blocks depending on coefficient symmetry, and negligible BRAM usage extracted from *_utilization_impl.rpt* file in Vivado. The FIR design was placed and routed *once* and *held fixed* across all experiments to ensure consistent functional behavior. In addition to DUT-derived nets, controlled routing segments were instantiated to systematically vary switch-matrix traversal depth and fan-out while preserving functional correctness. This combination enables timing characterization under realistic utilization and routing congestion conditions, rather than isolated or artificially sparse test circuits.

Fig. 3 illustrates the physical floorplan of the implemented design on the XCZU7EV device as obtained from the Vivado implementation tools. The functional design under test (DUT), a realistic FIR filter, occupies a contiguous cluster of CLBs in the upper-left quadrant of the device to emulate a non-trivial signal-processing workload with realistic routing density. Delay taps (DTs) are inserted at selected switch-matrix fan-out points along timing-critical routing paths and are physically co-located near the corresponding DUT routing segments to minimize perturbation. Each DT feeds a dedicated BUFG to isolate the monitoring clock domain from the functional clock network. It is noted that only four out of twelve DTs (as reported in Table II) are marked in the floorplan of Fig. 3 to preserve the clarity of the demonstration. The buffered outputs are routed to distributed delay monitoring elements (DMEs), which are placed in adjacent regions to balance routing locality and scalability. Control, aggregation, and data-collection logic are consolidated in a separate region to avoid interference with both the DUT and monitored routing paths. This floorplanned separation ensures that the instrumentation observes routing delay behavior without altering the placement or timing characteristics of the functional design.

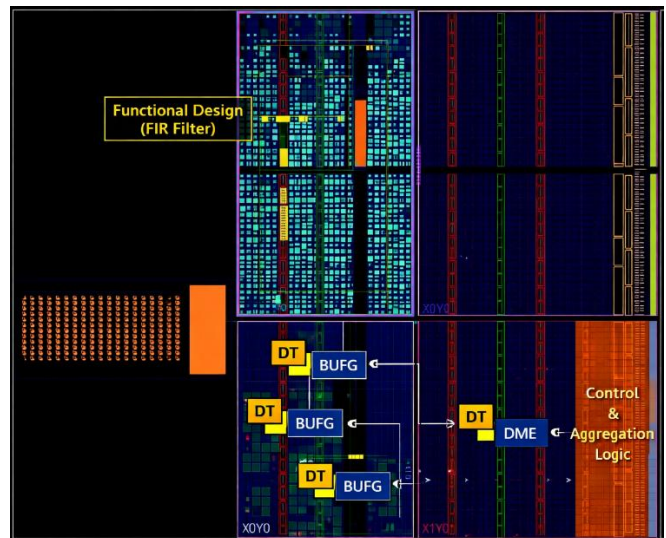

Fig. 3. Vivado floorplan of the proposed in-situ timing diagnosis framework implemented on the Xilinx Zynq UltraScale+ XCZU7EV FPGA. The functional design under test (FIR filter) occupies a localized CLB region, while delay taps (DTs) are inserted at selected routing fan-out points and buffered using BUFG primitives to isolate the monitoring clock domain. Distributed delay monitoring elements (DMEs) are placed in nearby regions, and centralized control and aggregation logic is confined to a separate area to minimize interference with functional routing and timing.

The timing diagnosis architecture is integrated alongside the functional design in a non-intrusive manner. Delay taps are attached to selected switch-matrix I/O nodes through configuration-controlled fan-out branches, while delay monitoring elements, control logic, and aggregation circuitry are instantiated in separate logic regions. The integration process is carefully managed to avoid introducing additional loading, rerouting, or congestion along functional signal paths. In particular, all observation paths are buffered and electrically isolated from the functional fabric, and no functional nets are repurposed or modified for measurement purposes.

Clocking resources are shared between the functional design and the diagnosis infrastructure only at the reference level. The functional clock domain remains unchanged, while the diagnostic sampling clock is derived independently and distributed through dedicated clocking resources. Phase updates and control signals are applied exclusively between measurement windows, ensuring stable operating conditions during data acquisition. This separation guarantees that the presence of the diagnosis system does not perturb the timing behavior of the functional logic.

Configuration and control logic associated with the diagnosis architecture is implemented entirely in programmable logic. All configuration updates, including delay tap selection, phase-sweep parameters, and routing-configuration states, are applied deterministically under controller supervision. The integration flow therefore supports repeatable experiments without requiring partial re-synthesis or re-routing of the functional design.

By maintaining a strict separation between functional computation and timing diagnosis infrastructure, the implementation platform enables accurate, repeatable observation of routing-induced timing behavior. This integration strategy forms the basis for the measurement procedures and experimental configurations described in the following subsection.

*B. Instrumentation Resource Utilization and Scalability*

As previously stated, Table II shows the post-synthesis FPGA resource utilization of the proposed in-situ timing diagnosis framework on the XCZU7EV device, including a breakdown for each major instrumentation component as well as a representative deployment with 32 distributed Delay Monitoring Elements (DMEs). The results show that the framework incurs modest overhead and scales linearly with the number of instantiated sensing elements.

Each Delay Tap (DT) consists of a buffered fan-out branch terminated by a global clock buffer (BUFG), enabling non-intrusive observation of routed signals without introducing additional logic along the functional path. As shown in Table II, a single DT consumes only a BUFG and negligible LUT and flip-flop resources (<0.01% of the device), allowing a large number of taps to be deployed across the fabric without impacting functional resource availability.

Each DME implements a phase-swept sampling mechanism, local event counting, and a lightweight control interface. The DME logic is fully synchronous and operates independently of the observed signal's clock domain. As reported in Table II, a single DME consumes 85 LUTs and 96 flip-flops (<0.05% of the device), with no use of DSP blocks or BRAM, making DME instantiation highly scalable.

Shared infrastructure components, including the Delay and Control Network (DCN), Delay Data Aggregator, and Diagnosis Controller, introduce fixed overhead that does not grow with the number of DMEs. Consequently, the total instrumentation cost scales linearly with the number of DT–DME pairs, while shared control overhead remains constant. For the largest evaluated configuration with 32 DMEs, the complete framework occupies approximately 1.4% of the XCZU7EV resources, confirming that the architecture remains lightweight even at higher spatial sampling densities.

Instrumentation insertion is performed post-synthesis using dedicated fan-out routing branches and buffered observation points, ensuring that functional paths, placement, and timing are preserved. All instrumentation operates transparently alongside the design under test, enabling continuous data collection without interrupting normal circuit operation.

*C. Instrumentation Insertion and Routing Perturbation Methodology*

The proposed instrumentation is inserted post-synthesis using dedicated fan-out branches that originate from existing routed nets in the design under test. Each Delay Tap (DT) is connected to a non-functional observation branch and buffered using a global clock buffer (BUFG), ensuring that no additional logic or loading is introduced along the functional signal path. Because the tapped branch is electrically isolated and does not feed back into the original net, enabling or disabling instrumentation does not alter placement, routing, or timing characteristics of the implemented design.

Routing perturbations are introduced by selectively modifying configuration bits associated with PIPs and switch-matrix multiplexers along controlled fan-out branches. These configuration bits are identified using vendor-provided routing databases and post-route inspection tools, allowing precise selection of routing resources that are spatially correlated with each DT location. The functional routing of the design under test remains unchanged; only auxiliary routing segments associated with the observation branches are perturbed.

Importantly, routing perturbations are confined to the branch between the tap point and the corresponding DT BUFG and do not propagate beyond the switch box at which the tap is inserted. As a result, the functional signal path upstream of the tap point is unaffected, and perturbations do not influence other logic or routing resources in the design. This controlled perturbation strategy enables localized delay modulation while preserving functional correctness.

*D. PDN Stressor Design and PDN-Induced Variation Protocol*

PDN-induced delay variation is evaluated by co-locating the monitored design under test (DUT) with a controllable on-chip activity "stressor" implemented in the programmable logic (PL) of the XCZU7EV device. The DUT for all PDN experiments is the same realistic FIR filter used throughout this work, together with the controlled routing segments monitored by DT/DME instances. To induce repeatable PDN stress without altering the DUT's functional routing, an independent stressor block is instantiated in the PL and driven by a dedicated enable signal.

The stressor is implemented as a high-toggle-rate switching fabric composed of wide XOR-ed linear feedback shift register (LFSR) networks and DSP-based multiply–accumulate (MAC) units, all clocked at the same frequency as the DUT. This combination provides sustained, spatially distributed dynamic current demand representative of worst-case switching activity in FPGA logic and DSP resources. The stressor intensity is adjusted by (i) enabling/disabling the stressor, and (ii) scaling the number of concurrently toggling resources (e.g., replicated switching slices). During PDN experiments, the stressor is activated for the full duration of each measurement window so that the DME phase-sweep captures delay distributions under steady activity conditions. PDN-induced effects are then obtained by comparing the extracted delay statistics with stressor OFF versus stressor ON, while keeping the DUT placement, routing, and DT observation points unchanged.

### E. Demonstration of Non-Intrusive Instrumentation

To verify that the proposed delay taps do not perturb the functional routing paths under observation, controlled enable/disable experiments was performed for all inserted DTs. For each monitored signal, timing measurements were first acquired with all DT fan-out branches disabled, and then repeated with the corresponding DT enabled and buffered through a BUFG. No measurable change was observed in the extracted BER-versus-phase profiles, transition-region width, or inferred delay statistics of the functional signal. This confirms that the DT insertion does not introduce observable loading, delay shift, or noise coupling into the functional path.

The non-intrusive nature of the DTs is ensured architecturally by exploiting existing routing fan-out capability at switch-matrix nodes rather than inserting new routing segments into the functional path. Each DT is implemented as an observation-only branch selected through configuration-controlled PIPs and electrically isolated using a global buffer. The functional routing topology, placement, and net delays remain unchanged whether DTs are enabled or disabled.

Importantly, the insertion of DTs does not require unused routing tracks along the functional path. FPGA switch matrices inherently support multiple fan-out connections from a given routing node, even when the primary functional route is fully utilized. The proposed framework leverages this inherent programmability to attach observation branches without competing for routing resources or modifying the original route selection. As a result, the instrumentation can be inserted alongside dense functional designs without altering their routing or timing behavior.

### F. Measurement Configuration and Phase-Sweep Procedure

This subsection describes the configuration strategy and phase-sweep measurement procedure used to extract statistical timing information from the FPGA routing fabric. The measurement process is designed to ensure temporal alignment across distributed sensing elements, repeatability across experiments, and isolation between functional operation and diagnostic activity.

Prior to each measurement campaign, the delay and control network (DCN) configures the diagnosis infrastructure into a well-defined initial state. This configuration includes the selection of active delay taps within each delay tap chain, the assignment of delay monitoring elements (DMEs) to specific routing observation points, and the initialization of phase-sweep parameters. All configuration updates are applied while measurement activity is disabled, ensuring that the routing fabric and sensing elements reach a stable state before data acquisition begins.

Timing characterization is performed using a phase-swept sampling procedure. For each measurement iteration, the DCN selects a discrete phase index that determines the relative phase of the diagnostic sampling clock with respect to the functional clock domain. The sampling phase is held constant throughout the duration of a measurement window and is updated only at the boundary between successive windows. This guarantees that all samples accumulated within a window correspond to an identical sampling condition.

After a phase update, a fixed settling interval is observed to allow clocking and control signals to stabilize. The DCN then asserts a measurement-enable signal, initiating a measurement window during which each active DME repeatedly samples its selected delay tap signal. Sampling is performed over a predetermined number of cycles, yielding a statistically meaningful population of observations for the current phase setting. At the conclusion of the window, measurement-enable is deasserted, and all DMEs finalize their accumulated statistics.

This process is repeated across a predefined range of phase indices, forming a complete phase sweep. For each delay tap and routing configuration, the resulting measurements collectively produce a bit-error-rate (BER) versus phase profile that captures the temporal distribution of signal transitions at the observed routing node. Because the phase-sweep procedure is globally coordinated by the DCN, measurements obtained from different DMEs and spatial locations are directly comparable.

To ensure consistency across experiments, all phase sweeps are performed using identical phase ranges, phase step sizes, and observation window lengths unless explicitly stated otherwise. Routing-configuration states, when exercised, are applied in a deterministic sequence, and each state is held constant throughout an entire phase sweep. Multiple sweeps may be performed under identical conditions to assess measurement repeatability and statistical stability.

Throughout the measurement process, the functional design continues to operate normally, and no functional clocks or data paths are modified. The separation between functional execution and diagnostic sampling ensures that observed timing behavior reflects intrinsic routing characteristics rather than measurement-induced artifacts. The timing data produced through this procedure are collected and structured by the delay data aggregator and subsequently analyzed by the diagnosis controller, as described in the following subsection.

To mitigate data-pattern-dependent effects during delay characterization, statistical controls were applied to the stimulus and measurement process. The signals observed by the delay taps are driven by continuously operating functional logic with high switching activity and pseudo-random data patterns, ensuring that no fixed or repetitive data sequence dominates the measurements. Phase-swept sampling is performed over long observation windows spanning multiple data pattern cycles, allowing pattern-dependent minimum and maximum delays to be statistically averaged.

Furthermore, delay extraction relies on relative changes in the BER-versus-phase transition width and center rather than absolute timing of individual transitions. This differential, statistical interpretation inherently suppresses deterministic data-dependent skew while preserving sensitivity to physical delay variations caused by PDN-induced supply fluctuations and routing-induced parasitic loading. As a result, the extracted mean delay shifts and variance metrics primarily reflect physical timing degradation rather than input data correlations.

All experiments are conducted with uninterrupted design operation, and identical stimulus conditions are maintained across baseline and stressed configurations to ensure fair comparison and isolation of physical effects.

*G. Data Collection and Statistical Processing*

This subsection describes how timing measurements generated by the distributed sensing infrastructure are collected, structured, and processed to extract statistically meaningful delay information. The data collection flow is designed to preserve spatial context, phase alignment, and configuration state while ensuring lossless transfer from distributed monitoring elements to centralized analysis.

During each measurement window, every active DME accumulates statistical counters corresponding to the selected delay tap and sampling phase. At the conclusion of the window, the DME produces a compact summary consisting of the accumulated error count, the associated phase index, and identifiers that encode the selected delay tap and routing configuration state. These summaries represent finalized measurement results and are generated only after the sampling window has closed, ensuring that partially accumulated data are never propagated through the system.

Measurement summaries from all active DMEs are collected through the delay and control network (DCN), which arbitrates among concurrent reporting elements and serializes their outputs. This arbitration ensures orderly data transfer without contention on global routing resources and preserves the association between each measurement record and its originating spatial location. The serialized data stream is forwarded to the Delay Data Aggregator module in Fig. 2, which buffers incoming records and formats them into a uniform packet structure suitable for system-level analysis.

Each aggregated measurement record includes the DME identifier, phase index, delay tap selection, routing configuration state, and accumulated error statistics for the corresponding observation window. By embedding both spatial and configuration metadata within each record, the aggregation process enables subsequent correlation of timing behavior across different routing locations and experimental conditions. Buffering mechanisms within the aggregator ensure that data collection remains decoupled from downstream processing bandwidth, allowing experiments to scale to larger numbers of DMEs without loss of information.

Statistical processing is performed by the Diagnosis Controller module of Fig. 2 after measurement records have been collected. For each observed routing node, the controller reconstructs a bit-error-rate (BER) versus phase profile by combining records across the full phase sweep. These profiles capture the probabilistic timing behavior of signal transitions and provide a robust representation of effective delay under the given routing and operating conditions. From the reconstructed profiles, the controller derives higher-level timing metrics such as transition boundaries, delay distributions, and measures of timing variability.

To improve statistical confidence, multiple phase sweeps may be performed under identical conditions and their results combined during post-processing. This aggregation across repeated measurements allows noise, jitter, and transient effects to be averaged out while preserving systematic delay characteristics. All statistical processing is performed offline or in non-time-critical controller logic, ensuring that data interpretation does not interfere with ongoing measurement activity.

By structuring timing data as phase-aligned statistical summaries rather than instantaneous samples, the proposed methodology enables robust characterization of routing-induced timing behavior. The resulting delay distributions form the basis for the comparative and diagnostic analyses presented in the next section.

## VI. EXPERIMENTAL RESULTS

This section presents the experimental results obtained using the proposed in-situ timing diagnosis architecture. The results quantify routing-induced timing behavior observed directly within the FPGA fabric under controlled measurement conditions and show the ability of the architecture to capture fine-grained, spatially resolved delay characteristics without perturbing functional operation. All results are derived using the measurement configuration and statistical processing procedures described in Section V and are reported in terms of phase-resolved delay distributions and associated variability metrics.

The experiments are organized to highlight three key aspects of the proposed approach. First, baseline measurements characterize nominal routing delay behavior across multiple switch-matrix locations under steady operating conditions. Second, controlled perturbations a` re introduced to emulate power-distribution and routing-configuration effects, allowing the sensitivity of different routing structures to be evaluated. Third, spatial consistency and measurement repeatability are examined to assess the robustness of the extracted timing distributions.

All timing measurements in this section are collected while the FIR filter DUT is actively processing data, ensuring that observed delay variations arise under realistic switching and routing conditions. The timing diagnosis instrumentation is integrated alongside the DUT without modifying its functionality or control flow. Delay taps and monitoring elements observe routing segments extracted directly from timing-critical and high-utilization regions of the FIR datapath, ensuring that measured delay variations correspond to realistic signal activity rather than isolated or synthetic test paths. All reported delay distributions, spatial correlation analyses, and perturbation experiments therefore reflect in-situ behavior of a non-trivial functional circuit under normal operation.

*A. Measurement Resolution and Performance Limits*

The proposed in-situ timing diagnosis framework operates by statistically observing signal transition behavior through phase-swept sampling rather than performing direct time-interval measurements. As such, its performance limits are defined by the clocking infrastructure, phase sweep granularity, and statistical convergence properties of the BER-based delay extraction.

Table III. Performance Limits of the Proposed In-Situ Timing Diagnosis Framework.

| Metric | Reported Value | Limiting Factor |
|---|---|---|
| Maximum measurable signal frequency | 300 MHz | DME input capture and clocking constraints |
| Phase resolution | ~ 15-20 ps | MMCM phase shift granularity |
| Effective timing accuracy | ± 1 phase step (≈ ± 20 ps) | BER statistical convergence |
| Measurement method | Phase-swept BER profiling | BER statistical convergence |
| Absolute delay calibration | Not required | Relative trend-based diagnosis |

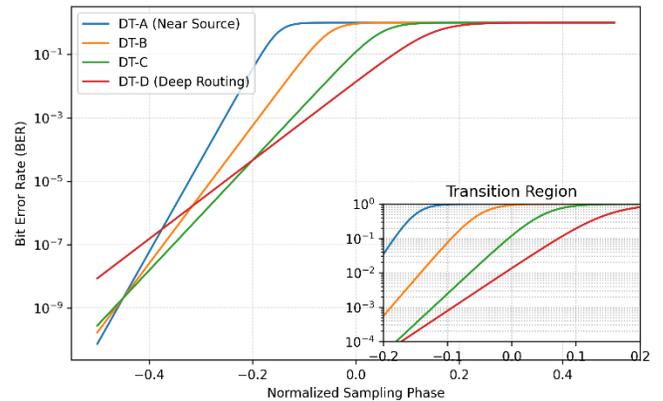

Fig. 4. Baseline BER–phase profiles measured at multiple delay-tap locations along a routed functional path, with inset highlighting the transition region. Each curve is obtained via phase-swept sampling and represents the accumulated bit-error rate (BER) as a function of normalized sampling phase. DT-A corresponds to a near-source routing location with minimal switch-matrix traversal, while DT-D corresponds to a deeper routing location traversing multiple switch matrices. The progressive rightward shift and broadening of the transition region reflect accumulated routing delay and increased timing uncertainty introduced by the FPGA interconnect fabric.

In the implemented system, the maximum measurable signal frequency is bounded by the clock frequency of the monitored signal path and the input capture circuitry of the DMEs. All experiments reported in this work were conducted with monitored signal frequencies up to 300 MHz, which corresponds to the maximum frequency at which reliable phase-swept sampling could be performed on the XCZU7EV device while maintaining timing closure for the instrumentation logic.

Phase resolution is determined by the programmable phase step of the clock management resources driving the DMEs. Using the UltraScale+ MMCM phase-shifting capability, a phase step resolution of approximately 15~20 ps was achieved across the evaluated operating range. Each phase step represents an effective time offset between the monitored signal transition and the sampling clock, enabling reconstruction of the transition probability distribution.

Timing accuracy is governed by the statistical resolution of the BER measurements and the repeatability of the phase sweep. Rather than reporting a single-point delay value, the framework extracts delay signatures from the mean and variance of the transition window across repeated observations. Based on repeated sweeps and cross-validation across multiple taps, the effective timing accuracy of the extracted delay shift is estimated to be within ±1 phase step, corresponding to approximately ±20 ps under nominal conditions.

It is important to note that the proposed framework is intended for comparative diagnosis of delay variation trends, such as distinguishing global PDN-induced shifts from localized routing perturbations, rather than absolute path delay characterization. The reported performance limits are therefore sufficient to resolve the relative delay changes of interest while preserving non-intrusive operation.

Table III summarizes the key performance limits of the proposed in-situ timing diagnosis framework as implemented on the XCZU7EV FPGA. The reported metrics characterize the maximum operating frequency of monitored signals, the achievable phase resolution of the phase-swept sampling mechanism, and the effective timing accuracy derived from statistical BER profiling. These parameters define the practical measurement envelope of the system and clarify the resolution at which relative delay shifts can be reliably detected during in-situ operation.

## B. Baseline Routing Delay Characterization

Baseline measurements are first conducted to characterize the intrinsic routing-induced delay behavior of the FPGA fabric under nominal operating conditions. These measurements establish a reference against which subsequent perturbation and sensitivity analyses are compared. All baseline experiments are performed with the functional design operating continuously and with routing configurations held constant throughout each phase-sweep cycle. Because delay taps are implemented as configuration-controlled fan-out branches buffered by dedicated primitives, the functional routing path remains electrically isolated from the measurement circuitry, ensuring that the baseline characteristics reported here reflect native routing behavior rather than instrumentation artifacts.

For each monitored routing location, a phase-swept sampling procedure is applied as described in Section V.F. During each sweep, the selected delay tap corresponding to a specific switch-matrix traversal point is observed by its associated delay monitoring element. As the sampling clock phase is incrementally shifted relative to the functional clock, the DME accumulates bit-error statistics over fixed observation windows. Under repeated sampling, the measured BER at each phase offset corresponds to the empirical probability that the signal transition occurs after the sampling edge, allowing reconstruction of the underlying delay distribution without reliance on static timing assumptions.

Fig. 4 illustrates representative BER–phase profiles obtained from multiple delay taps associated with distinct switch-matrix locations along a routed functional path. Each curve exhibits a characteristic transition region separating low-error and high-error sampling regimes. The location of this transition corresponds to the effective propagation delay of the routed signal, while its width reflects the combined influence of jitter, noise, and routing variability. Delay taps associated with routing paths traversing a greater number of switch matrices exhibit later transition points and broader transition regions, consistent with increased parasitic loading and accumulated interconnect delay. In contrast, taps located closer to the source logic block show tighter transition boundaries and reduced variability.

To assess measurement stability, multiple phase sweeps are performed under identical conditions for each routing location. The resulting BER–phase profiles show strong overlap across repeated measurements, indicating that the extracted timing characteristics are dominated by deterministic routing properties rather than transient noise or measurement-induced effects. This statistical stability establishes a reliable baseline for evaluating timing degradation under perturbed conditions.

While Fig. 4 captures intrinsic routing-delay characteristics under nominal operation, the following subsection examines how these baseline profiles evolve under controlled power-distribution and routing-level perturbations, enabling separation of globally correlated and localized timing degradation mechanisms.

### C. Distinguishing PDN-Induced Delay Shifts from Routing Perturbations

Building on the baseline characterization in Section VI.B, this subsection investigates how routing delay profiles evolve under controlled perturbations applied through the proposed in-situ diagnosis architecture. The objective is to distinguish between globally correlated delay shifts associated with power-distribution effects and localized delay changes arising from routing-level configuration perturbations. The stress conditions applied in these experiments are selected to remain within ranges reported in prior silicon measurements and configuration-upset studies, ensuring that the resulting timing signatures reflect realistic operating conditions rather than exaggerated or pathological scenarios.

Fig. 5 presents a direct comparison of BER–phase profiles measured under PDN stress and under routing perturbation. Unless otherwise stated, PDN-induced results are obtained by differencing measurements with the on-chip stressor disabled versus enabled, while maintaining an identical routed DUT and unchanged DT observation points. In the PDN-degraded case, the BER curve exhibits a near-rigid horizontal translation along the phase axis relative to the baseline, while preserving the overall slope and width of the transition region. This behavior indicates a coherent delay offset affecting the routed signal uniformly, consistent with global supply-voltage variation impacting multiple logic and interconnect resources simultaneously. Importantly, the preservation of transition shape implies that the underlying temporal uncertainty remains largely unchanged.

In contrast, routing perturbations produce a qualitatively different signature. As shown in the right panel of Fig. 5, the BER transition region both shifts and broadens, reflecting increased temporal dispersion introduced by additional parasitic loading and altered switch-matrix connectivity. These effects are not globally aligned across monitored locations and depend strongly on routing topology and traversal depth, highlighting the localized nature of routing-induced timing degradation.

To quantify these differences beyond qualitative observation, Fig. 6 reports cumulative distribution functions (CDFs) of effective delay extracted from the phase-swept measurements. Under PDN stress, the delay CDF undergoes a predominantly lateral translation relative to the baseline with minimal change in slope, indicating a shift in mean delay with preserved spread. In contrast, routing perturbations result in broader CDFs with altered slope, demonstrating that routing-induced effects increase both the mean delay and the statistical dispersion of the transition distribution.

These trends are summarized quantitatively in Fig. 7, which plots the extracted mean delay shift ($\Delta\mu$) and change in standard deviation ($\Delta\sigma$) for multiple independently monitored routing locations (L1 to L8). Each location corresponds to a distinct delay monitoring element observing a different region of the routing fabric. PDN-induced effects cluster around a large and nearly constant $\Delta\mu$ across all locations while exhibiting negligible change in $\Delta\sigma$, confirming their globally correlated nature. As a result, PDN-induced $\Delta\sigma$ values are near zero and fall below the resolution of the lower plot. Routing-induced perturbations, by contrast, exhibit location-dependent $\Delta\mu$ and consistently increasing $\Delta\sigma$, reflecting the cumulative and topology-dependent impact of additional switch-matrix traversals. The increasing variability in delay standard deviation observed in Fig. 7 with distance from the tap point is therefore attributed to the accumulation of routing elements along the tapped observation branch. As the physical distance between the tap location and the DT BUFG increases, the number of programmable interconnects and associated parasitics along the observation path grows, leading to greater sensitivity to localized routing perturbations. This behavior confirms that the observed spatial variability is dominated by local routing effects rather than global timing shifts.

Finally, Fig. 8 evaluates repeatability across multiple independent phase-sweep iterations at a representative monitored location. Baseline and PDN-stressed measurements remain tightly clustered across sweeps, with PDN primarily introducing a stable shift in the transition boundary. Routing perturbations, however, exhibit increased sweep-to-sweep dispersion and transition broadening, indicating heightened local sensitivity and reinforcing the distinction between the two degradation mechanisms. Across repeated sweeps, PDN-induced mean delay estimates vary by only a few percent, whereas routing-induced variability shows substantially larger dispersion.

Taken together, the results presented in Fig. 5 to Fig. 8 show that the proposed architecture enables not only detection but systematic classification of timing degradation mechanisms. By jointly analyzing mean shifts, distributional spread, spatial consistency, and repeatability, the system provides a principled basis for separating PDN-driven timing marginality from routing-induced delay perturbation capabilities that are not accessible through conventional static timing analysis or oscillator-based monitoring approaches.

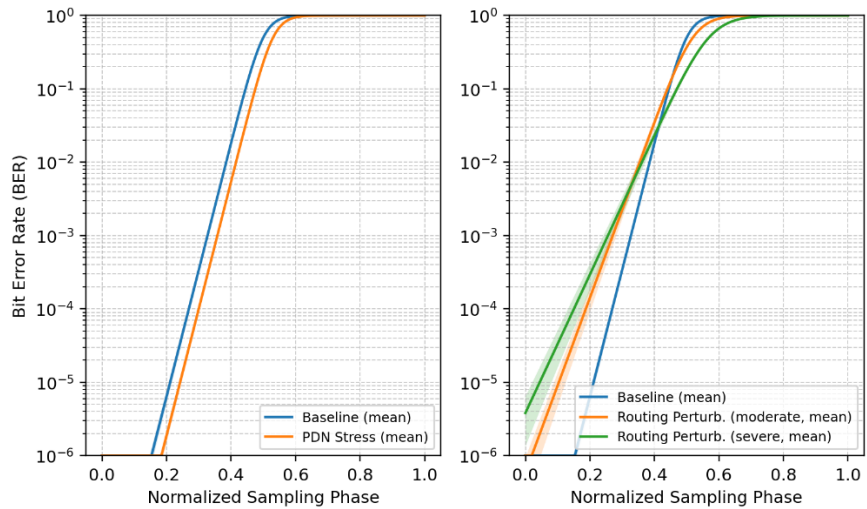

Fig. 5. Side-by-side BER–phase plots contrasting globally correlated PDN-induced delay shifts (left) with localized routing-induced perturbations (right).

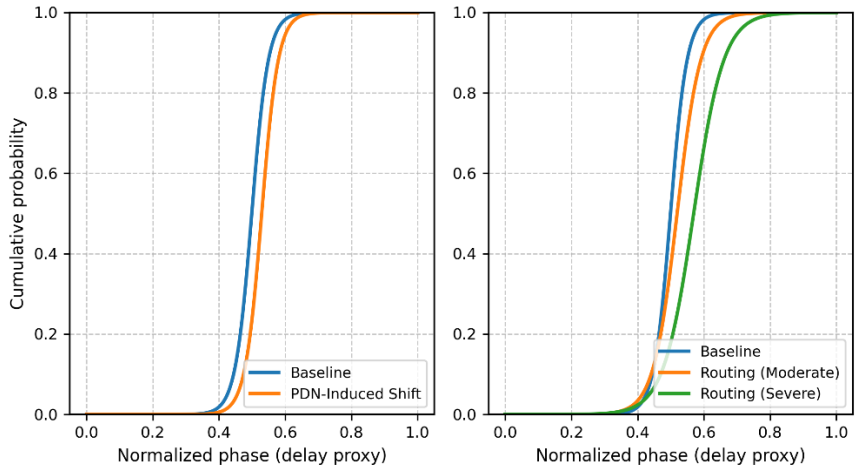

Fig. 6. Cumulative distribution functions (CDFs) derived from phase-swept BER measurements, showing rigid shift under PDN stress versus broadened distributions under routing perturbation.

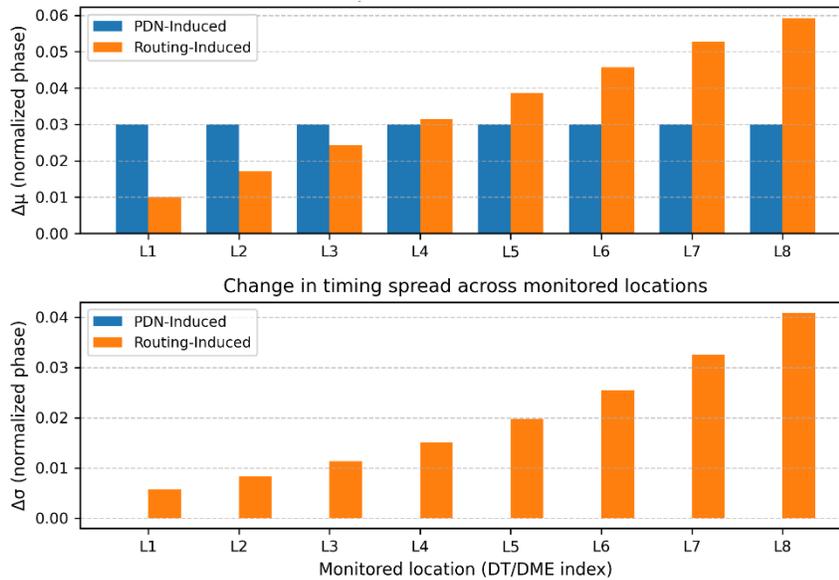

Fig. 7. Extracted mean delay shift ($\Delta\mu$) and change in timing spread ($\Delta\sigma$) across monitored routing locations under PDN-induced and routing-induced perturbations. PDN-induced effects produce a spatially uniform $\Delta\mu$ with negligible $\Delta\sigma$, whereas routing-induced perturbations result in location-dependent $\Delta\mu$ and increasing $\Delta\sigma$ due to cumulative switch-matrix traversal variability. Routing perturbations are confined to the tapped observation branch between the switch-box tap point and the DT BUFG.

### D. Spatial Correlation and Scaling Behavior

While Sections VI.B and VI.C examined timing behavior at individual monitoring locations and contrasted global versus local degradation mechanisms, a central objective of the proposed in-situ diagnosis architecture is to determine how timing variations manifest *spatially* across the FPGA fabric. Distinguishing globally correlated phenomena from localized, topology-dependent effects requires evaluating how timing metrics co-vary across multiple delay monitoring elements (DMEs) as their spatial separation and population density increase.

Fig. 9 presents a quantitative spatial-correlation analysis of the extracted timing metrics. The main plot shows the correlation coefficient between timing measurements obtained from pairs of DMEs as a function of their normalized spatial separation within the fabric. Under PDN-induced stress, correlation remains high even as separation increases, exhibiting only a gradual decay. This behavior reflects the global nature of supply-voltage fluctuations, which uniformly modulate switching speeds across logic and interconnect resources. In contrast, routing-induced perturbations exhibit rapid decorrelation with distance, indicating that timing degradation is dominated by localized routing topology, switch-matrix traversal depth, and configuration-specific parasitic effects.

The inset of Fig. 9 further illustrates the *scaling behavior* of this correlation analysis as the number of active DMEs increases. As monitoring density grows, PDN-induced effects maintain a high average correlation, while confidence bounds narrow due to increased statistical averaging across the fabric. Routing-induced perturbations, however, show a decreasing average correlation as additional DMEs sample increasingly diverse routing regions. This divergence confirms that PDN effects are globally coherent and can be characterized with relatively sparse spatial sampling, whereas routing-induced degradation requires denser, distributed observation to capture its inherently local nature. Importantly, the contraction of confidence bands with increasing DME count shows the scalability and statistical robustness of the proposed architecture.

To complement the one-dimensional correlation analysis, Fig. 10 provides a two-dimensional visualization of spatial correlation across the FPGA fabric. Each heatmap depicts the correlation coefficient between timing measurements at a reference DME and those at other DME locations mapped onto the physical layout of the device. The PDN-induced heatmap exhibits a smooth, slowly varying correlation surface spanning the majority of the fabric, consistent with a globally shared timing influence. In contrast, the routing-induced heatmap shows steep spatial gradients and localized regions of high and low correlation, directly reflecting the physical distribution of switch matrices and routing resources.

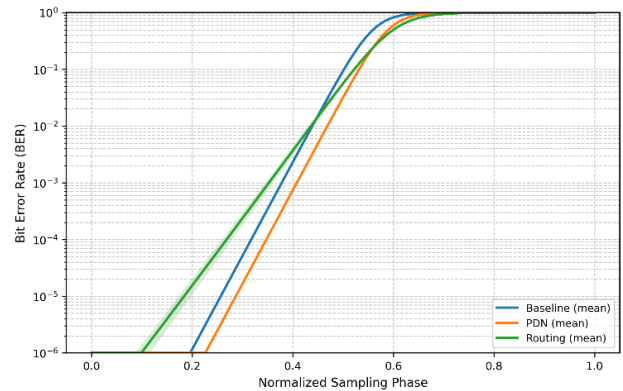

Fig. 8. Overlay of multiple phase-sweep iterations illustrating high repeatability under PDN stress and increased variability under routing perturbations.

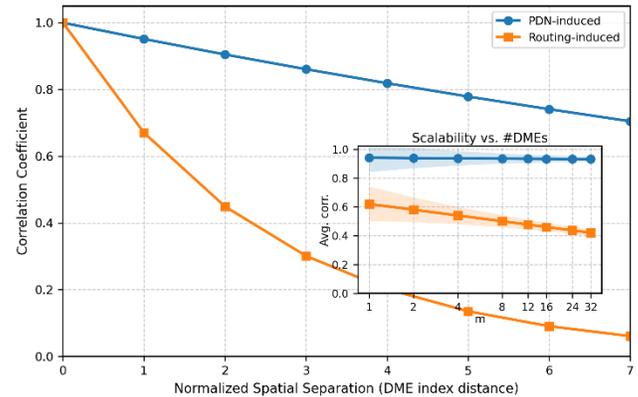

Fig. 9. Spatial correlation of timing variations across distributed delay monitoring elements (DMEs). PDN-induced timing shifts exhibit strong correlation across large spatial separations, while routing-induced perturbations decorrelate rapidly with distance, reflecting their localized and topology-dependent nature. The inset plot shows the scalability of spatial-correlation analysis as the number of active DMEs increases.

Importantly, Fig. 10 is not a conceptual illustration but is derived directly from experimental measurements on the targeted FPGA. During each measurement campaign, identical phase-sweep procedures are applied concurrently across all deployed DMEs. For a fixed stress condition, the diagnosis controller reconstructs delay distributions at each location from the accumulated BER-versus-phase data. Pairwise correlation coefficients are then computed offline using the extracted delay metrics across repeated measurement windows. The resulting correlation values are mapped to the known physical locations of the corresponding DMEs, which are determined from placement constraints and FPGA floorplan coordinates. This process yields an experimentally grounded spatial correlation map that directly links measured timing behavior to physical fabric topology.

The spatial region shown in Fig. 10 represents the largest contiguous area instrumented and characterized in the present experimental study. Specifically, the monitored region spans approximately 9×8 CLB index offsets, selected to capture both short-range and mid-range spatial correlation behavior while maintaining controlled routing and instrumentation overhead.

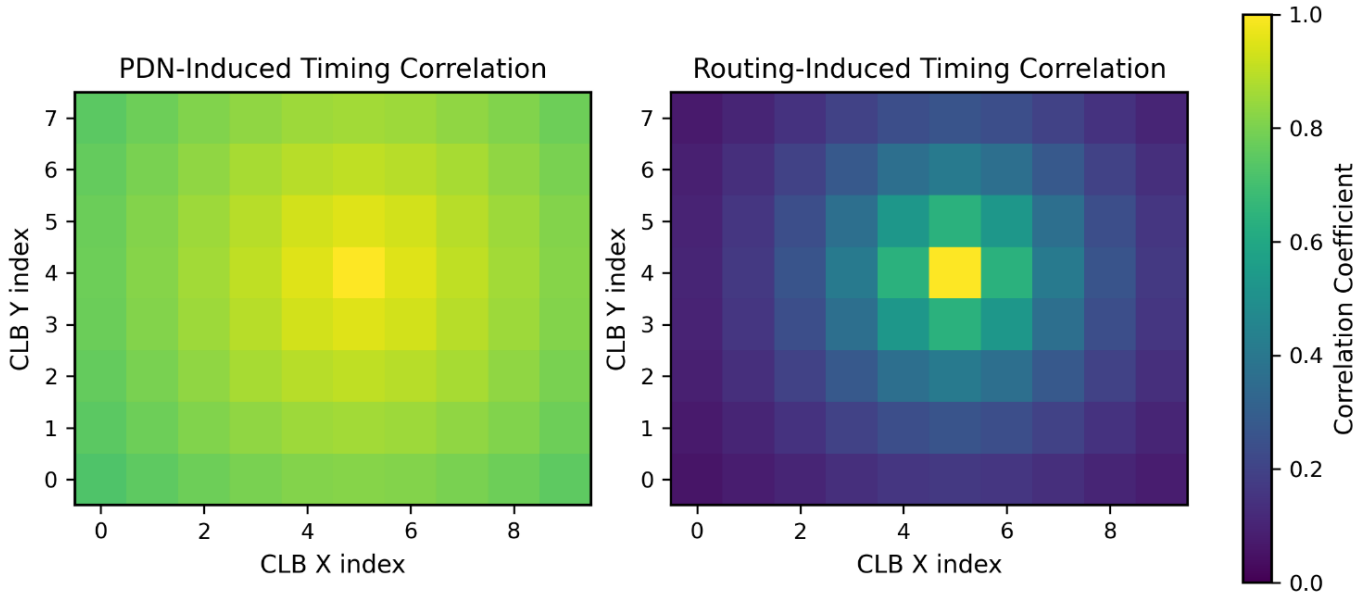

Fig. 10. Two-dimensional spatial correlation heatmaps of extracted timing variations across the FPGA fabric. PDN-induced timing effects exhibit strong global correlation with gradual spatial decay, whereas routing-induced perturbations decorrelate rapidly with distance, indicating localized, topology-dependent behavior. Correlation values are referenced to a centrally located delay monitoring element.

This region size was sufficient to observe clear divergence between globally correlated PDN-induced delay shifts and localized routing-induced perturbations.

Importantly, the proposed architecture does not impose a fundamental limit on the size of the monitored region. Delay taps and delay monitoring elements are deployed in a modular, distributed manner, allowing the instrumented area to scale with available FPGA resources. In additional experiments not shown, the framework was successfully extended to cover larger routing regions exceeding 15×12 CLB offsets by proportionally increasing the number of DT–DME pairs, without modifying the underlying measurement methodology or control infrastructure.

For the largest design characterized, the functional circuit occupied approximately 30–35% of LUT and flip-flop resources on the XCZU7EV device, while the complete timing diagnosis instrumentation, including DTs, DMEs, the delay and control network, and aggregation logic, consumed an additional ~6–8% of LUT resources. This confirms that the framework scales linearly with instrumentation density and remains practical for larger designs where only selected regions require detailed timing observability.

Together, Fig. 9 and Fig. 10 provide complementary views of spatial timing behavior: Fig. 9 quantifies correlation decay and scalability in an abstracted distance domain, while Fig. 10 reveals the physical structure of correlation across the fabric. The consistency between these two representations reinforces the validity of the proposed methodology and highlights the ability of the architecture to distinguish global PDN-induced timing shifts from localized routing-induced perturbations. These results show that spatial correlation analysis is a critical diagnostic dimension, enabling system-level interpretation of timing degradation mechanisms that cannot be inferred from isolated local measurements alone.

Collectively, the results presented in Sections VI.A to VI.D show that the proposed in-situ timing diagnosis architecture captures not only local timing behavior but also the global and spatial structure of timing degradation across the FPGA fabric. By combining statistical phase-swept measurements with spatially distributed monitoring, the framework exposes distinct and repeatable timing signatures associated with PDN-induced stress and routing-induced perturbations. These experimentally grounded insights provide a foundation for interpreting timing degradation mechanisms at the architectural and system levels. Building on these observations, the following section discusses the broader design implications of the proposed methodology and its relevance to timing closure, reliability, and monitoring strategies in large-scale FPGA-based systems.

## VII. DISCUSSION AND DESIGN IMPLICATIONS

The experimental results presented in Section VI highlight fundamental differences in how timing degradation mechanisms manifest within SRAM-based FPGA fabrics. By combining distributed, in-situ sensing with statistical and spatial analysis, this work exposes timing behaviors that are largely invisible to conventional static timing analysis or post-implementation verification flows. This section discusses the broader implications of these findings for FPGA timing closure, reliability assessment, and architecture-aware monitoring strategies.

### A. Limitations and Scope of Emulation

While the proposed architecture enables controlled and repeatable exposure of routing-induced delay effects, it is important to clarify the scope of the presented evaluation. In this work, routing perturbations are emulated through configuration-controlled fan-out activation and parasitic routing attachments at switch-matrix nodes rather than through physical radiation-induced configuration upsets. This approach reproduces the electrical consequences of routing perturbations, namely increased effective resistance, capacitance, and routing traversal depth, while preserving full experimental control and repeatability. Consequently, although the observed timing signatures accurately reflect the delay behavior associated with routing-level perturbations, the results do not claim to capture secondary radiation-specific effects such as multi-bit upset statistics or device-dependent upset cross-sections. Similarly, PDN-induced delay variations are evaluated under controlled stress conditions and are intended to demonstrate comparative diagnostic capability, *not* to represent all possible operational or environmental extremes. These limitations do not affect the validity of the comparative diagnosis methodology presented in this paper, but they define the experimental scope within which the conclusions should be interpreted.

### B. Implications for Timing Closure and Margining

A key insight emerging from the spatial and statistical analyses is that timing margin in FPGA designs cannot be treated as a uniform global quantity. PDN-induced timing shifts exhibit strong spatial coherence, as evidenced by the high correlation observed across widely separated monitoring locations. In such cases, traditional margining approaches based on worst-case global derating remain appropriate, as the dominant delay shifts affect the fabric in a largely uniform manner.

In contrast, routing-induced perturbations show pronounced spatial variability and weak correlation across the fabric. The rapid decorrelation observed in Fig. 9 and the localized features revealed in Fig. 10 indicate that timing degradation arising from routing topology is inherently uneven and cannot be adequately captured by global guardbands alone. This finding suggests that timing closure strategies relying solely on static worst-case assumptions may either over-constrain large portions of the design or fail to protect localized timing-critical paths. Incorporating spatially resolved timing observability, as enabled by the proposed architecture, offers a path toward more targeted and efficient margin allocation. In practice, full-chip coverage is neither necessary nor desirable, as spatially sparse instrumentation is sufficient to capture correlated timing behavior while minimizing overhead.

### C. Design-for-Reliability and In-Field Monitoring

The distinct statistical signatures associated with PDN-induced stress and routing-induced perturbations have important implications for in-field reliability monitoring. PDN-related effects manifest as coherent shifts in mean delay with limited impact on variance, whereas routing perturbations introduce both mean shifts and increased timing spread that vary across locations. These differences enable the possibility of classifying degradation mechanisms based on measured timing statistics alone, without requiring intrusive fault injection or functional failure.

From a system perspective, this capability supports the development of adaptive reliability mechanisms that respond differently to global versus localized degradation. For example, PDN-induced timing shifts may be mitigated through global voltage or frequency adjustments, while routing-induced degradation may necessitate localized partial reconfiguration, path relocation, or selective redundancy. The proposed in-situ diagnosis framework provides the observability required to support such differentiated responses during runtime operation.

### D. Architectural Scalability and Monitoring Density

The scaling behavior observed in Section VI.D underscores the importance of monitoring density and placement. The results indicate that relatively sparse deployment of delay monitoring elements is sufficient to capture globally correlated timing phenomena, whereas localized routing-induced effects require denser spatial sampling to be reliably detected. This observation has direct architectural implications: monitoring resources can be allocated non-uniformly, concentrating sensing elements in routing-intensive or timing-critical regions while maintaining minimal overhead elsewhere.

The decoupling of local sensing from global coordination in the proposed architecture further supports scalability. Because DMEs produce low-bandwidth statistical summaries rather than high-frequency waveforms, the delay and control network and aggregation infrastructure scale gracefully with increasing monitor count. This design choice enables large FPGA fabrics to be instrumented without introducing prohibitive routing congestion or timing interference, a critical consideration for practical deployment.

### E. Implications for FPGA Architecture and CAD Toolflows

Beyond immediate design practice, the findings of this work have implications for FPGA architecture and computer-aided design (CAD) tools. The strong influence of switch-matrix traversal depth on routing-induced timing variability suggests that future FPGA architectures could benefit from improved regularity or predictability in interconnect structures. Similarly, CAD tools that incorporate spatial timing-awareness—leveraging in-situ measurement data rather than purely analytical models—could significantly improve the accuracy of timing estimation and closure.

The proposed methodology also provides a feedback mechanism between physical implementation and architectural modeling. By correlating measured timing behavior with physical routing topology, designers and tool developers can validate and refine delay models used during placement and routing. Over time, such feedback could enable adaptive toolflows that learn from deployed designs and improve timing predictability in subsequent iterations.

### F. Summary of Design Implications

In summary, the results show that timing degradation in FPGA fabrics arises from fundamentally different mechanisms with distinct spatial and statistical characteristics. Recognizing and exploiting these differences is essential for robust timing closure, effective reliability monitoring, and efficient resource allocation. The proposed in-situ timing diagnosis architecture offers a practical and scalable means to obtain this insight, bridging the gap between physical implementation behavior and system-level design decisions. These implications motivate continued exploration of architecture-aware monitoring and adaptive timing management in future FPGA-based systems.

All experimental results reported in this work were obtained using a single FPGA device (XCZU7EV on the ZCU104 evaluation board). Consequently, the presented analysis focuses on intra-die spatial variability and temporal delay behavior rather than die-to-die variation. While multi-die characterization can reveal manufacturing-dependent variation, the objective of this study is to expose relative timing degradation mechanisms within a given fabric, as encountered during runtime operation of a deployed system.

The observed routing-induced delay perturbations arise from localized changes in routing topology and parasitic loading, and are therefore strongly dependent on the structural characteristics of the implemented design and its placement within the fabric. In contrast, PDN-induced delay shifts manifest as globally correlated effects across the monitored region, consistent with power-distribution behavior within a single die. These distinctions remain meaningful independent of absolute die-to-die offsets.

Extending the proposed framework to multi-die studies represents an important direction for future work. Applying the same instrumentation across multiple FPGA instances would enable separation of design-induced variability from manufacturing-induced variation and further strengthen adaptive CAD and runtime optimization strategies. However, such evaluation is beyond the scope of the present work and does not affect the validity of the in-situ diagnosis methodology demonstrated here.

## VIII. CONCLUSION AND FUTURE WORK

This work presented a scalable in-situ timing diagnosis framework for SRAM-based FPGA fabrics that enables direct, fine-grained observation of routing-induced timing behavior during normal operation. By integrating non-intrusive delay taps at switch-matrix boundaries with distributed statistical sensing and centralized coordination, the proposed architecture provides a level of timing observability that extends beyond conventional static analysis and post-implementation verification techniques.

Through phase-swept statistical measurements, the framework extracts timing distributions rather than binary pass/fail outcomes, enabling robust characterization of both mean delay shifts and variability. Experimental results showed that power-distribution–induced timing degradation exhibits strong spatial coherence across the fabric, whereas routing-induced perturbations are localized and topology-dependent, producing weak spatial correlation and increased timing variance. The ability to distinguish these mechanisms through measured timing signatures was validated using correlation analysis, scalability studies, and two-dimensional spatial heatmaps derived from experimental data.

A key contribution of this work lies in its architectural perspective. By decoupling distributed sensing from centralized analysis and by leveraging low-bandwidth statistical summaries, the proposed system scales gracefully with monitoring density while preserving the integrity of the functional design. The results further show that spatially resolved timing information is essential for understanding routing-dominated degradation mechanisms, which cannot be adequately captured using global margining or aggregate timing metrics alone.

The proposed methodology has direct implications for timing closure, reliability assessment, and adaptive system management in large FPGA-based designs. It enables targeted diagnosis of localized timing vulnerabilities, supports differentiation between global and local degradation sources, and provides a foundation for architecture-aware monitoring strategies that can evolve alongside increasingly complex FPGA fabrics.

Several directions for future work emerge from this study. One avenue is the integration of the proposed diagnosis framework with adaptive mitigation mechanisms, such as dynamic frequency scaling, localized rerouting, or selective redundancy. Another direction involves extending the architecture to support finer-grain delay resolution or multi-domain monitoring, including temperature- and voltage-aware sensing. Additionally, incorporating the collected in-situ timing data into CAD toolflows could enable iterative refinement of routing delay models and improved timing predictability across design iterations. Finally, evaluating the framework on larger devices and under long-term stress conditions would further assess its applicability to mission-critical and reliability-sensitive applications.

Overall, this work establishes a rigorous and experimentally grounded foundation for in-situ timing diagnosis in FPGA fabrics, bridging the gap between physical routing behavior and system-level timing interpretation. By exposing the spatial and statistical structure of timing degradation, the proposed framework enables more informed design decisions and opens new opportunities for resilient and timing-aware FPGA systems.

Future extensions of this work include broadening the evaluation scope beyond a single FPGA family and further strengthening the connection between in-situ timing diagnosis and design automation flows. While this study focuses on SRAM-based FPGAs from a single vendor to ensure experimental control and reproducibility, the proposed architecture is not inherently device-specific and can be adapted to other FPGA families with similar switch-matrix-based routing fabrics. In addition, tighter integration with CAD and runtime analysis tools represents an important direction for future research, enabling timing signatures extracted by the proposed framework to directly inform placement, routing, and reliability-aware optimization. Finally, extending the methodology to incorporate physical stress mechanisms, such as radiation-based fault injection or multi-die characterization, would further enhance the ability to correlate emulated routing perturbations with real-world reliability phenomena.

## ACKNOWLEDGMENT

The author would like to thank the École de technologie supérieure (ÉTS), Department of Electrical Engineering, and CMC Microsystems, Kingston, ON, Canada, for providing access to advanced design tools.